\documentclass[conference]{IEEEtran}
\IEEEoverridecommandlockouts
\usepackage[colorlinks,linkcolor=blue,anchorcolor=blue,citecolor=blue]{hyperref}
\usepackage{cite}
\usepackage{url}
\usepackage[caption=false]{subfig}
\usepackage{float}
\usepackage{makecell}
\usepackage{amsmath,amssymb,amsfonts}
\hypersetup{hypertexnames=false}
\allowdisplaybreaks[4]
\usepackage{graphicx}
\usepackage{enumerate}
\usepackage{array}
\usepackage{multirow}
\usepackage{color}
\usepackage{colortbl}
\usepackage[ruled,vlined,linesnumbered]{algorithm2e}
\usepackage{tikz}

\newcommand{\Tset}{\mathcal{T}}
\newcommand{\Gset}{\mathcal{G}}
\newcommand{\Nset}{\mathcal{N}}
\newcommand{\Lset}{\mathcal{L}}

\def\BibTeX{{\rm B\kern-.05em{\sc i\kern-.025em b}\kern-.08em
    T\kern-.1667em\lower.7ex\hbox{E}\kern-.125emX}}
\begin{document}

\title{Voltage Ride-Through-Constrained Power Capping for Large-Scale Data Centers\\

\thanks{This work was supported by the National Key Research and Development Program under Grant 2024YFB4206500.}
}

\author{\IEEEauthorblockN{Jie Zhu, Ye Tian, and Yinliang Xu*}
\IEEEauthorblockA{\textit{Shenzhen International Graduate School} \\
\textit{Tsinghua University}\\
Shenzhen, China \\
zhuj24@mails.tsinghua.edu.cn, tiany25@mails.tsinghua.edu.cn, xu.yinliang@sz.tsinghua.edu.cn}

}

\maketitle
\setlength{\abovedisplayskip}{5.5pt plus 1pt minus 1pt}
\setlength{\belowdisplayskip}{5.5pt plus 1pt minus 1pt}
\setlength{\abovedisplayshortskip}{0pt plus 1pt}
\setlength{\belowdisplayshortskip}{5.5pt plus 1pt minus 1pt}

\begin{abstract}
Large-scale data centers may satisfy steady-state network constraints while violating operational limits during voltage disturbances. This paper proposes a voltage ride-through (VRT)-aware power-capping framework that integrates AC optimal power flow scheduling with dynamic ride-through assessment. The proposed dynamic model captures converter limits, DC-link energy dynamics, battery support, and IT-bus voltage response. A simulation-based checker evaluates VRT constraints under prescribed faults, and the power caps are iteratively updated to obtain feasible schedules. Case studies on the IEEE 14-bus system with two data centers demonstrate that steady-state feasible operating points may violate transient voltage and power-import limits. Selective power capping achieves approximately 5\% demand reduction during capped periods while satisfying the prescribed VRT requirements. The results highlight the necessity of incorporating dynamic security assessment into data-center operational limit determination.
\end{abstract}

\section{Introduction}
The rapid growth of artificial intelligence (AI) applications has accelerated the deployment of large-scale data centers (DCs), leading to rapidly increasing electricity demand and new challenges for power system operation \cite{sun2025energy}. Unlike conventional passive loads, modern data centers are characterized by high-density computing infrastructures and power-electronic interfaces \cite{azizi2025grid}. Consequently, large-scale DCs are becoming increasingly coupled with power system operation and require enhanced operational flexibility and security considerations.

To enhance the flexibility and sustainability of data-center operation, recent studies have explored energy management and demand response strategies that coordinate computing workloads with grid conditions and operational requirements \cite{11367141}. For example, the flexibility of Internet data centers has been utilized to enable workload migration according to electricity price signals \cite{sang2026coordinating} and support grid services through participation in ancillary service markets \cite{11023622}. Furthermore, carbon-aware and emission-aware operation strategies have been developed to reduce operational costs and environmental impacts by coordinating computing workloads with renewable energy availability \cite{radovanovic2023carbon}.

Among various demand-side management approaches, power capping has emerged as an effective mechanism for regulating data-center electricity consumption without requiring detailed workload-level coordination \cite{11197299}. By setting an upper bound on the allowable power consumption, power capping can reshape the load profile according to electricity prices, carbon intensity, or other grid signals. Existing studies have explored power capping from different perspectives, including application-aware power allocation, computing performance management, and adaptive energy optimization \cite{sun2025power}.

However, existing power capping methods mainly determine operational limits based on economic objectives, workload characteristics, or computing service constraints, while assuming that the obtained power trajectories are feasible from the perspective of power system dynamic security. In particular, the transient behavior of large-scale data centers during grid disturbances has received limited consideration in operational scheduling. Recent studies \cite{ren2026hosting,xie2026data} have shown that voltage disturbances may trigger uninterruptible power supply (UPS) control actions, converter current limitations, and battery energy storage system (BESS)-supported operation, leading to significant changes in grid-side power exchange. Therefore, a power trajectory that satisfies steady-state network constraints may still result in insufficient ride-through capability or excessive disturbance-induced load reduction.

To address this gap, this paper proposes a dynamic-security-aware power capping framework for large-scale data centers under voltage ride-through (VRT) constraints. The proposed framework integrates steady-state AC optimal power flow scheduling with grid dynamic simulations and an internal data-center response model, where VRT violations are identified to iteratively update secure power limits. The dynamic model considers IT workload, cooling and auxiliary demand, DC-link energy dynamics, BESS support, and UPS converter constraints. Simulation results demonstrate that steady-state feasible schedules may violate VRT requirements, while the proposed dynamic-constrained power capping strategy maintains ride-through capability with limited power reduction.

\section{VRT-Secure Data-Center Power Capping}
For a given data center configuration, the power-cap trajectory defines the maximum allowable active power consumption in the day-ahead scheduling stage:
\begin{equation}
P^{\mathrm{cap}}=[P_1^{\mathrm{cap}},P_2^{\mathrm{cap}},\ldots,P_T^{\mathrm{cap}}]
\label{eq:dc_cap_trajectory}
\end{equation}
where $P_t^{\mathrm{cap}}$ denotes the power cap at period $t$. The actual data-center power consumption is determined by
\begin{equation}
P_t^{\mathrm{DC}}=\min(P_t^{\mathrm{req}},P_t^{\mathrm{cap}})
\label{eq:actual_dc_power}
\end{equation}
where $P_t^{\mathrm{req}}$ represents the requested power.
\subsection{Steady-State Optimal Scheduling Model}
The day-ahead scheduling problem jointly minimizes the generation cost and the cost associated with DC power capping. The resulting ACOPF problem is formulated as
\begin{subequations}
\label{eq:day_ahead_acopf}
\begin{align}
&\min\sum_{t\in\Tset}\left[\sum_{i\in\Gset}C_i(P_{i,t}^{G})+\sum_{i\in\mathcal N_{\mathrm{DC}}}c_{i,t}^{\mathrm{cap}}\left(P_{i,t}^{\mathrm{req}}-P_{i,t}^{\mathrm{cap}}\right)\right]
\label{eq:objective}\\
&\text{s.t.}\quad 0\leq P_{i,t}^{\mathrm{cap}}\leq P_{i,t}^{\mathrm{req}}
\label{eq:dc_cap_bound}\\
&\begin{aligned}[t]
P_{i,t}^{G}-P_{i,t}^{L}-\chi_iP_{i,t}^{\mathrm{cap}}&
={}V_{i,t}\sum_{j\in\Nset}V_{j,t}\\
&\times\left[G_{ij}\cos\theta_{ij,t}+B_{ij}\sin\theta_{ij,t}\right]
\end{aligned}
\label{eq:acpf_p}\\
&\begin{aligned}[t]
Q_{i,t}^{G}-Q_{i,t}^{L}-\chi_iQ_{i,t}^{\mathrm{DC}}&
={}V_{i,t}\sum_{j\in\Nset}V_{j,t}\\
&\times\left[G_{ij}\sin\theta_{ij,t}-B_{ij}\cos\theta_{ij,t}\right]
\end{aligned}
\label{eq:acpf_q}\\
&Q_{i,t}^{\mathrm{DC}}=P_{i,t}^{\mathrm{cap}}\tan\left(\cos^{-1}(\mathrm{pf}_i)\right)
\label{eq:dc_q}\\
&\begin{aligned}[t]
P_{ij,t}
=&V_{i,t}V_{j,t}\left[G_{ij}\cos\theta_{ij,t}+B_{ij}\sin\theta_{ij,t}\right]-G_{ij}V_{i,t}^{2}
\end{aligned}
\label{eq:branch_pflow}\\
&\begin{aligned}[t]
Q_{ij,t}
=&B_{ij}V_{i,t}^{2}+V_{i,t}V_{j,t}[G_{ij}\sin\theta_{ij,t}\!-\!B_{ij}\cos\theta_{ij,t}]
\end{aligned}
\label{eq:branch_qflow}\\
&|S_{ij,t}|^{2}=P_{ij,t}^{2}+Q_{ij,t}^{2}, ~ |S_{ij,t}|\leq\overline S_{ij}
\label{eq:branch_sflow}\\
&\underline P_i\leq P_{i,t}^{G}\leq\overline P_i,~ \underline Q_i\leq Q_{i,t}^{G}\leq\overline Q_i,~ \underline V_i\leq V_{i,t}\leq\overline V_i
\label{eq:gen_limits}\\
&-R_i^{\mathrm{down}}\leq P_{i,t}^{G}-P_{i,t-1}^{G}\leq R_i^{\mathrm{up}}
\label{eq:generator_ramping}
\end{align}
\end{subequations}
where $t\in\Tset$, $i,j\in\Nset$, $(i,j)\in\Lset$, $i\in\Gset$ for generator-related quantities, and $i\in\mathcal N_{\mathrm{DC}}$ for data-center-related quantities. $C_i(\cdot)$ and $c_{i,t}^{\mathrm{cap}}$ denote the generation cost and capping penalty, respectively. $P_{i,t}^{G}$, $Q_{i,t}^{G}$, $P_{i,t}^{L}$, and $Q_{i,t}^{L}$ denote generation and conventional load; $\chi_i$ indicates whether a data center is connected to bus $i$; $Q_{i,t}^{\mathrm{DC}}$ and $\mathrm{pf}_i$ denote its reactive-power demand and power factor. $V_{i,t}$ and $\theta_{i,t}$ are the bus-voltage magnitude and phase angle, with $\theta_{ij,t}=\theta_{i,t}-\theta_{j,t}$; $P_{ij,t}$, $Q_{ij,t}$, and $S_{ij,t}$ are the branch active, reactive, and apparent powers; and $G_{ij}$ and $B_{ij}$ are the corresponding admittance terms. $\underline P_i$, $\overline P_i$, $\underline Q_i$, $\overline Q_i$, $R_i^{\mathrm{up}}$, $R_i^{\mathrm{down}}$, $\underline V_i$, $\overline V_i$, and $\overline S_{ij}$ denote the generator, ramping, voltage, and line-flow limits.

\subsection{Pre-Fault Operating Point and Dynamic Fault Simulation}
For a given scheduling solution, the ACOPF determines the pre-fault operating point as
\begin{equation}
x_t^0=[P_t^G,Q_t^G,V_t,\theta_t,P_t^{\mathrm{cap}}]
\label{eq:prefault_point}
\end{equation}
For each contingency $f\in\mathcal F$, the grid dynamic simulation is performed based on the pre-fault operating point, and the resulting voltage trajectory at the data-center point of interconnection (POI) is obtained as
\begin{equation}
V_{t,f}^{\mathrm{POI}}(\tau)=G_{\mathrm{grid}}(x_t^0,f)
\label{eq:vpoi}
\end{equation}
where $G_{\mathrm{grid}}(\cdot)$ denotes the grid dynamic simulation operator and $\tau$ represents the post-disturbance time.

\subsection{Data-Center Internal Dynamic Model}
\label{subsec:dc_model}
\begin{figure}[!t]
\centering
\includegraphics[width=0.47\textwidth]{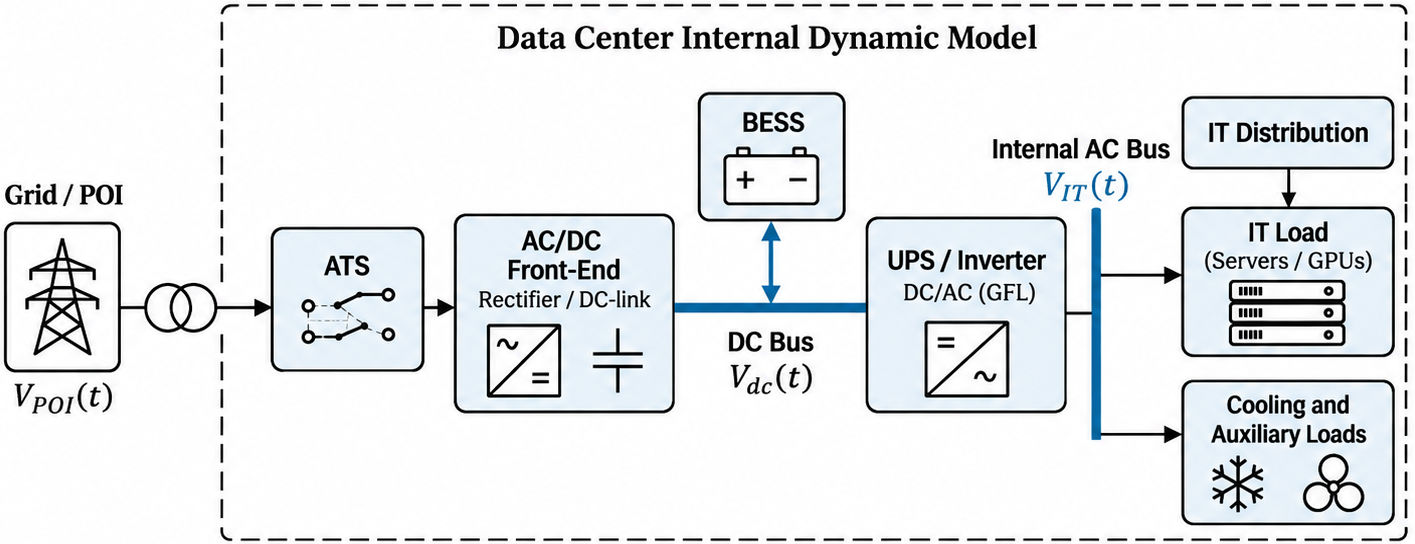}
\caption{Data-center internal dynamic model architecture.}
\label{fig:dynamic_model}
\end{figure}

The DC dynamic model, as illustrated in Fig.~\ref{fig:dynamic_model}, is driven by the fault-dependent POI voltage waveform $V_{t,f}^{\mathrm{POI}}(\tau)$ obtained from the grid dynamic simulation. The internal DC system model comprises the IT workload, cooling and auxiliary power demands, DC-link energy storage, a DC-side BESS, and the grid-connected UPS converter.

\subsubsection{IT, Cooling, and Auxiliary Loads}
The scheduled power cap $P^{\mathrm{cap}}$ represents the maximum allowable facility active-power demand, including the IT, cooling, and auxiliary loads:
\begin{equation}
P^{\mathrm{cap}}=P^{\mathrm{IT}}+P^{\mathrm{cool}}+P^{\mathrm{aux}}
\label{eq:dc_cap_components}
\end{equation}
where $P^{\mathrm{IT}}$ is the IT load, $P^{\mathrm{cool}}$ is the cooling load, $P^{\mathrm{aux}}$ is the auxiliary loads.

To assess the most stressed operating condition associated with a given power cap, the data center is assumed to operate at its scheduled maximum demand during the disturbance:
\begin{equation}
P_t^{\mathrm{DC}}(\tau)=P_t^{\mathrm{cap}}.
\label{eq:dc_dynamic_load}
\end{equation}

\subsubsection{DC-Link Energy Dynamics}
The DC link provides short-term energy buffering between the grid-side UPS converter, the DC-side BESS, and the protected loads. The stored energy is represented as
\begin{equation}
E^{\mathrm{dc}}(\tau)=\frac12 C_{\mathrm{dc}}(V^{\mathrm{dc}}(\tau))^2
\label{eq:dc_energy}
\end{equation}
where $C_{\mathrm{dc}}$ and $V^{\mathrm{dc}}$ denote the DC-link capacitance and voltage, respectively.

The DC-link energy balance is given by
\begin{equation}
\dot E^{\mathrm{dc}}=P^g+P^{\mathrm{BESS}}-P^{\mathrm{DC}}
\label{eq:dc_balance}
\end{equation}
where $P^g$ is the grid-side active power supplied by the UPS converter, $P^{\mathrm{BESS}}$ denotes the BESS output power, and $P^{\mathrm{DC}}$ represents the protected DC-side demand.

The IT-bus voltage response is represented by a first-order approximation:
\begin{equation}
T_{\mathrm{IT}}\frac{dV^{\mathrm{IT}}(\tau)}{d\tau}+V^{\mathrm{IT}}(\tau)=\frac{V^{\mathrm{dc}}(\tau)}{V^{\mathrm{dc,nom}}}
\label{eq:it_voltage_response}
\end{equation}
where $V^{\mathrm{IT}}$ denotes the normalized IT-bus voltage, $V^{\mathrm{dc,nom}}$ is the nominal DC-link voltage, and $T_{\mathrm{IT}}$ is the converter response time constant. The initial condition is set as $V^{\mathrm{IT}}(0)=1$.

\subsubsection{UPS Supervisory Logic}
The UPS supervisory controller determines the operating mode according to the POI voltage magnitude and the corresponding dwell-time logic. Four voltage-dependent operating regions are considered: normal grid operation, mixed operation, LVRT transient operation, and full battery-supported operation. The supervisory state is represented as
\begin{equation}
\sigma(\tau)=\mathcal S_{\mathrm{UPS}}(|V^{\mathrm{POI}}(\tau)|,T_{\mathrm{low}}),\quad \sigma\in\{\mathsf N,\mathsf M,\mathsf L,\mathsf B\}
\label{eq:ups_supervisor}
\end{equation}
where $\mathsf N$, $\mathsf M$, $\mathsf L$, and $\mathsf B$ denote normal grid operation, mixed operation, LVRT transient operation, and full battery-supported operation, respectively. The variable $T_{\mathrm{low}}$ denotes the continuous duration of a severe voltage depression.

The voltage-region-based supervisory logic is defined as
\begin{equation}
\sigma(\tau)=
\begin{cases}
\mathsf N,& |V^{\mathrm{POI}}(\tau)|\geq V_{\mathrm N},\\[1mm]
\mathsf M,& V_{\mathrm B}\leq |V^{\mathrm{POI}}(\tau)|<V_{\mathrm N},\\[1mm]
\mathsf L,& |V^{\mathrm{POI}}(\tau)|<V_{\mathrm B},\ T_{\mathrm{low}}\leq T_{\mathrm B},\\[1mm]
\mathsf B,& |V^{\mathrm{POI}}(\tau)|<V_{\mathrm B},\ T_{\mathrm{low}}>T_{\mathrm B}.
\end{cases}
\label{eq:ups_voltage_region}
\end{equation}
where $V_{\mathrm N}$ and $V_{\mathrm B}$ denote the voltage thresholds between normal/mixed operation and mixed/LVRT operation, respectively. $T_{\mathrm B}$ is the dwell time required to activate full battery-supported operation.

In normal mode, the BESS regulates any residual DC-link energy deviation to restore the DC-link energy reserve for potential subsequent voltage dips. The BESS returns to idle operation as the energy deviation vanishes. During mixed operation, the grid-side converter supplies the available power limited by its voltage-dependent capability, while the BESS compensates for the remaining power deficit. During LVRT transient operation, the BESS supports the protected IT load without transferring the facility into full battery operation. When the severe voltage depression persists beyond $T_{\mathrm B}$, the UPS enters full battery-supported operation and the grid-side power exchange is significantly reduced.

\subsubsection{DC-Side BESS Model}
During voltage disturbances, the BESS provides DC-side power support when the grid-side UPS converter cannot fully supply the protected load demand. The normalized DC-link energy deviation and the grid--load power mismatch are defined as
\begin{equation}
e^{\mathrm{dc}}=\frac{E^{\mathrm{dc,nom}}-E^{\mathrm{dc}}}{E^{\mathrm{dc,nom}}},\quad \Delta P=P^{\mathrm{DC}}-P^g
\label{eq:energy_power_errors}
\end{equation}
where $E^{\mathrm{dc}}$ and $E^{\mathrm{dc,nom}}$ denote the actual and nominal DC-link energies, respectively. $P^{\mathrm{DC}}$ represents the DC-side demand, and $P^g$ denotes the grid-side UPS converter power output. A positive value of $\Delta P$ indicates insufficient grid-side power support and requires additional BESS discharge power.

The BESS power command $P^{\mathrm{BESS},\star}$ is determined by the UPS operating state:
\begin{equation}
P^{\mathrm{BESS},\star}=
\begin{cases}
\mathrm{sat}(K_{p,e}e^{\mathrm{dc}}),&\sigma=\mathsf N\\[1mm]
\mathrm{sat}(K_{p,e}e^{\mathrm{dc}}+K_{p,P}\Delta P),&\sigma=\mathsf M\\[1mm]
\mathrm{sat}(\Delta P+K_{p,e}e^{\mathrm{dc}}),&\sigma\in\{\mathsf L,\mathsf B\}
\end{cases}
\label{eq:bess_control_logic}
\end{equation}
where positive $P^{\mathrm{BESS}}$ denotes discharging and negative values represent charging. The gain $K_{p,e}$ regulates the DC-link energy deviation, while $K_{p,P}$ provides instantaneous power-mismatch compensation during voltage disturbances. The saturation operator $\mathrm{sat}(\cdot)$ limits the commanded power within the allowable BESS charging and discharging bounds:
\begin{equation}
-\overline P^{\mathrm{BESS,ch}}\leq P^{\mathrm{BESS}}\leq\overline P^{\mathrm{BESS,dis}}.
\label{eq:bess_power_bounds}
\end{equation}
The actual BESS power follows the commanded reference through the converter response dynamics:
\begin{equation}
\tau_{\mathrm{BESS}}\dot P^{\mathrm{BESS}}=P^{\mathrm{BESS},\star}-P^{\mathrm{BESS}}
\label{eq:bess_power_response}
\end{equation}
where $\tau_{\mathrm{BESS}}$ denotes the BESS power response time constant.

The BESS energy dynamics are described as
\begin{equation}
\dot E^{\mathrm{BESS}}=
\begin{cases}
-\dfrac{P^{\mathrm{BESS}}}{\eta_{\mathrm{dis}}},&P^{\mathrm{BESS}}\geq0\\[1mm]
-\eta_{\mathrm{ch}}P^{\mathrm{BESS}},&P^{\mathrm{BESS}}<0
\end{cases}
\label{eq:bess_energy}
\end{equation}
subject to
\begin{equation}
\underline E^{\mathrm{BESS}}\leq E^{\mathrm{BESS}}\leq\overline E^{\mathrm{BESS}}.
\label{eq:bess_energy_bounds}
\end{equation}
This study considers fault ride-through and short-term post-fault recovery; subsequent BESS recharging is outside its scope. The energy limits are checked throughout the evaluation horizon using the actual BESS power. A candidate power cap is rejected if either limit is violated.

\subsubsection{Grid-Side UPS Converter Capability}
The converter capability determines the maximum active and reactive power exchange during voltage disturbances. The converter current constraint is
\begin{equation}
|I(\tau)|\leq I^{\max}
\label{eq:converter_current_limit}
\end{equation}
which yields
\begin{equation}
(P^g)^2+(Q^g)^2\leq(S_B|V^{\mathrm{POI}}|I^{\max})^2
\label{eq:converter_power_limit}
\end{equation}
where $S_B$ is the converter base power and $I^{\max}$ is the normalized current limit determined by the installed UPS rating.

Assuming active-power-priority operation, when grid-side power transfer is enabled, the actual active power is
\begin{equation}
P^g=\min\left\{P^{g,\star},\sqrt{(S_B|V^{\mathrm{POI}}|I^{\max})^2-(Q^g)^2}\right\}
\label{eq:ups_active_power}
\end{equation}
where $P^{g,\star}=P^{\mathrm{DC}}$ is the nominal grid-side power reference required to supply the DC-side demand. The UPS supervisory state overrides this reference during battery-supported and reconnection operation.

\subsection{VRT Security Criteria}
The VRT requirement is evaluated from two aspects: the serviceability of the protected IT load and the disturbance-induced POI load reduction. The first criterion ensures that the internal IT system maintains an acceptable voltage level during the disturbance, while the second criterion limits excessive customer-initiated load loss observed from the power system.

For each period $t$ and disturbance scenario $f$, the sustained IT-bus voltage is defined based on the minimum voltage within a specified evaluation window $\Delta T$:
\begin{equation}
V_{\mathrm{IT},\Delta}^{t,f}=\min_\tau\left\{\min_{s\in[\tau,\tau+\Delta T]}V_{t,f}^{\mathrm{IT}}(s)\right\}
\label{eq:sustained_it_voltage}
\end{equation}
where $V_{\mathrm{IT},\Delta}^{t,f}$ denotes the worst-case sustained protected IT-bus voltage during disturbance $f$ at period $t$. Let $V_{\mathrm{IT}}^{\mathrm{req}}$ denote the minimum acceptable IT-bus voltage required to maintain protected IT service. The internal voltage ride-through margin is defined as
\begin{equation}
M^{V,t,f}=V_{\mathrm{IT},\Delta}^{t,f}-V_{\mathrm{IT}}^{\mathrm{req}}
\label{eq:voltage_margin}
\end{equation}
where a nonnegative margin indicates that the protected IT load remains serviceable during the disturbance.

In addition to the internal voltage criterion, the disturbance-induced POI load reduction is considered to limit excessive customer-initiated load loss. The maximum active-power reduction is defined as
\begin{equation}
L_{t,f}^{\mathrm{loss}}=\max_\tau\{0,P_t^{\mathrm{POI},0}-P_{t,f}^{\mathrm{POI}}(\tau)\}
\label{eq:poi_load_loss}
\end{equation}
where $P_t^{\mathrm{POI},0}$ denotes the pre-disturbance POI active-power import and $P_{t,f}^{\mathrm{POI}}(\tau)$ denotes the transient POI active-power exchange during disturbance $f$.

The allowable POI load reduction is represented by
\begin{equation}
L_{t,f}^{\mathrm{loss}}\leq\lambda_{\mathrm{loss}}P_t^{\mathrm{POI},0}
\label{eq:poi_load_loss_limit}
\end{equation}
where $\lambda_{\mathrm{loss}}$ denotes the maximum allowable fraction of customer-initiated POI load reduction.

Therefore, a power-cap trajectory is considered VRT-secure if both the internal IT voltage and POI load reduction requirements are satisfied:
\begin{equation}
M^{V,t,f}\geq0,\quad L_{t,f}^{\mathrm{loss}}\leq\lambda_{\mathrm{loss}}P_t^{\mathrm{POI},0},\quad\forall t,f.
\label{eq:vrt_security}
\end{equation}

\subsection{Iterative VRT-Secure Scheduling Framework}
Embedding the complete fault simulation and facility-level VRT dynamics directly into the ACOPF \eqref{eq:day_ahead_acopf} results in a computationally intensive dynamic optimization problem. Therefore, an iterative master--checker framework is adopted.

At iteration $k$, the master problem solves
\begin{equation}
z^{(k)}=\arg\min_z J(z)
\label{eq:iterative_master}
\end{equation}
subject to the operating constraints in \eqref{eq:day_ahead_acopf} and the current VRT-secure power limits $\overline P_{i,t}^{(k)}$:
\begin{equation}
P_{i,t}^{\mathrm{cap}}\leq\overline P_{i,t}^{(k)},\quad i\in\mathcal N_{\mathrm{DC}},\quad t\in\mathcal T.
\label{eq:iterative_cap}
\end{equation}
The obtained pre-fault operating points are subsequently evaluated using the grid dynamic model and the data-center VRT model under all disturbance scenarios $f\in\mathcal F$. For a data center $i$ at period $t$ that violates any VRT criterion, a local bisection search is performed to determine the power limit $\overline P_{i,t}^{(k)}$. The search interval is initialized as $P_{i,t}^{\mathrm L}\leq P_{i,t}^{\mathrm{cap}}\leq P_{i,t}^{\mathrm U}$. At each iteration, the trial power cap is set to the midpoint $P_{i,t}^{\mathrm M}=(P_{i,t}^{\mathrm L}+P_{i,t}^{\mathrm U})/2$, and the corresponding operating point is assessed against the VRT constraints. If all considered contingencies satisfy the VRT requirements, the lower bound is updated to $P_{i,t}^{\mathrm M}$; otherwise, the upper bound is updated to $P_{i,t}^{\mathrm M}$. The bisection search terminates when $P_{i,t}^{\mathrm U}-P_{i,t}^{\mathrm L}\leq\epsilon_P$, and the resulting power limit is adopted as $\overline P_{i,t}^{(k)}$.
\section{Case Study}
\begin{figure}[!t]
\centering
\includegraphics[width=0.38\textwidth]{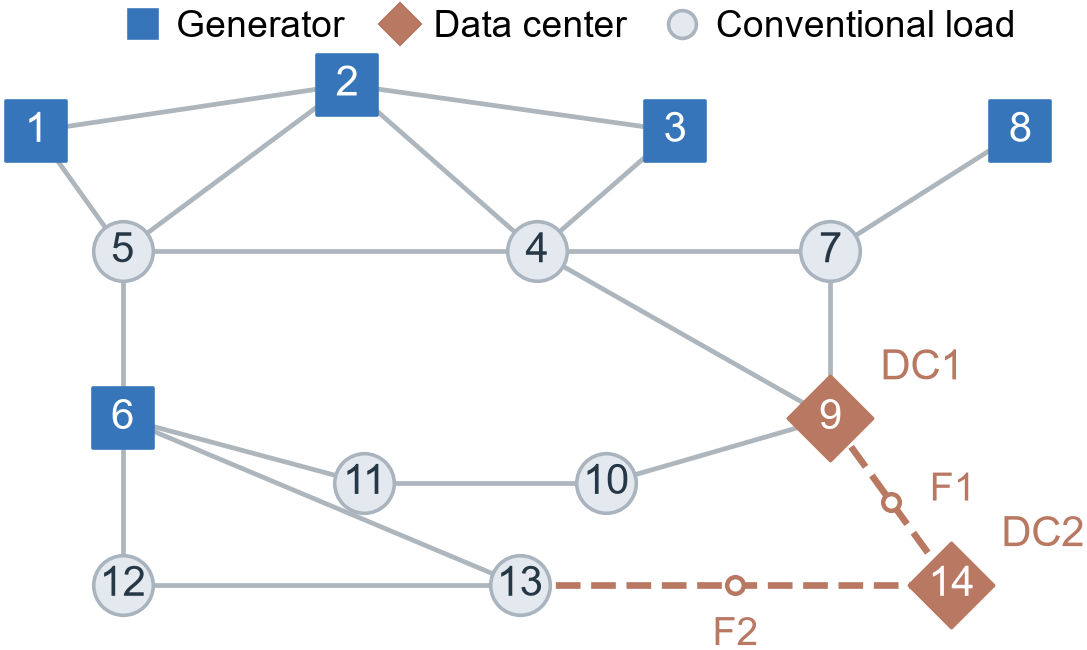}
\caption{IEEE 14-bus test system and fault locations F1 and F2.}
\label{fig:topology}
\end{figure}

The proposed framework is evaluated on the IEEE 14-bus test system shown in Fig.~\ref{fig:topology}. Two large-scale data centers, denoted as DC1 and DC2, are connected to buses 9 and 14, respectively. Demand profiles representing sustained AI-training workloads are considered, with daily mean requested powers of 75.10~MW for DC1 and 56.28~MW for DC2. The ACOPF subproblems are solved using IPOPT.

Two fault scenarios, denoted as F1 and F2, are considered. F1 represents a three-phase short-circuit fault at the midpoint of line 9--14, whereas F2 represents the same type of fault at the midpoint of line 13--14. Each fault is initiated at $t=0.30$~s and cleared after 150~ms. Grid transient simulations are performed using ANDES, and the resulting POI voltage trajectories are provided as inputs to the data-center dynamic models. The IT-bus voltage is required to remain above 0.9~pu, while the transient reduction in grid-side active-power import is limited to 20\% of its pre-fault value.

\begin{figure}[!t]
\centering
\includegraphics[width=0.45\textwidth]{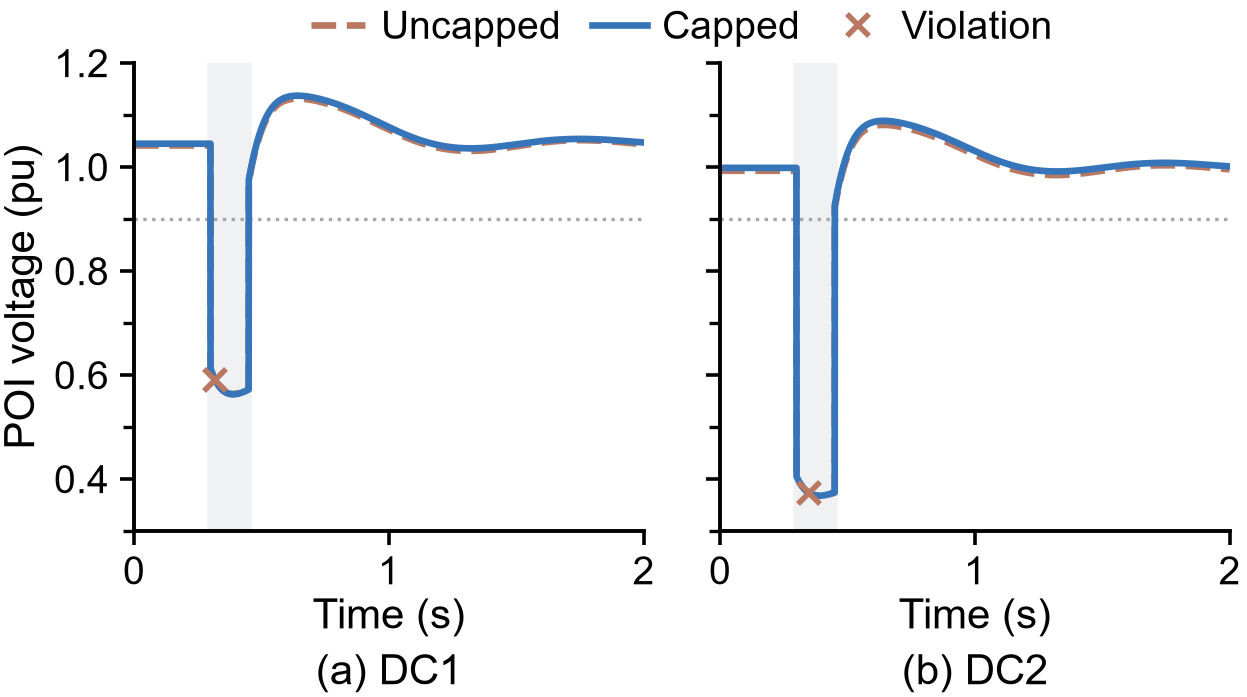}
\caption{POI voltage responses under fault F1.}
\label{fig:fig3_v_poi_F1}
\end{figure}

Without power capping, both data centers satisfy the ride-through criteria under fault F2 throughout the scheduling horizon. Therefore, the following transient analysis focuses on fault F1, where the combined requested power reaches its maximum. Fig.~\ref{fig:fig3_v_poi_F1} presents the corresponding POI voltage responses. Both data centers experience voltage sags, with a more severe voltage depression observed at DC2. Although the POI voltages recover after fault clearance, the transient internal voltage response and grid-side power capability determine the ride-through performance.

\begin{figure}[!t]
\centering
\includegraphics[width=0.47\textwidth]{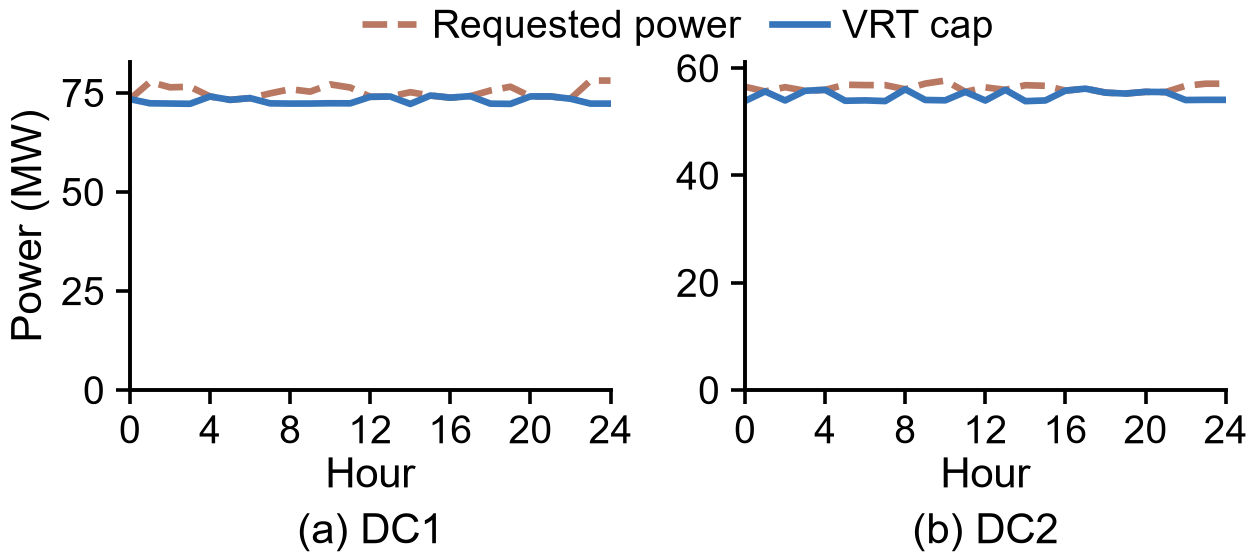}
\caption{Requested power and VRT-constrained power caps.}
\label{fig:fig7_daily_caps}
\end{figure}

Fig.~\ref{fig:fig7_daily_caps} compares the requested powers with the VRT-constrained power caps. Power capping is activated in 12 out of the 24 scheduling periods for each data center, while the requested powers are maintained in the remaining periods. The average reductions during capped periods are 5.24\% for DC1 and 5.05\% for DC2. These scheduled reductions adjust the pre-fault operating points to accommodate the limited power-transfer capability during voltage sags.

\begin{figure}[!t]
\centering
\includegraphics[width=0.47\textwidth]{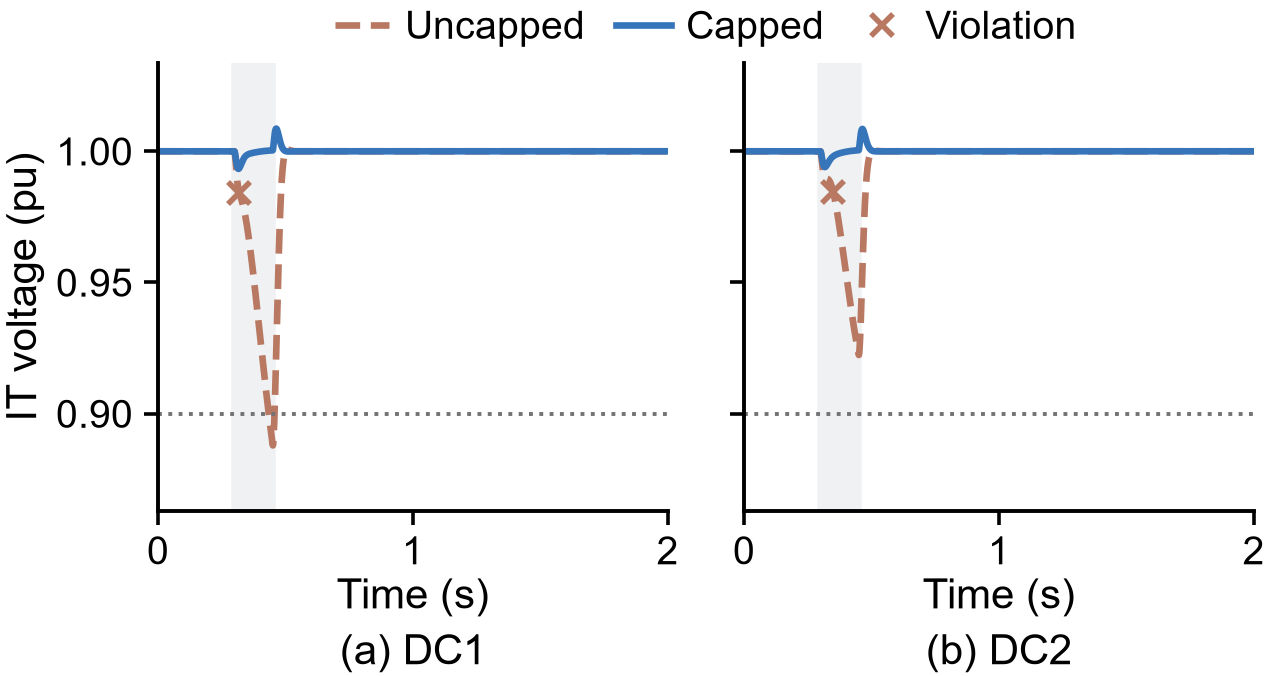}
\caption{IT-bus voltage responses under fault F1.}
\label{fig:fig4_v_it_F1}
\end{figure}

\begin{figure}[!t]
\centering
\includegraphics[width=0.45\textwidth]{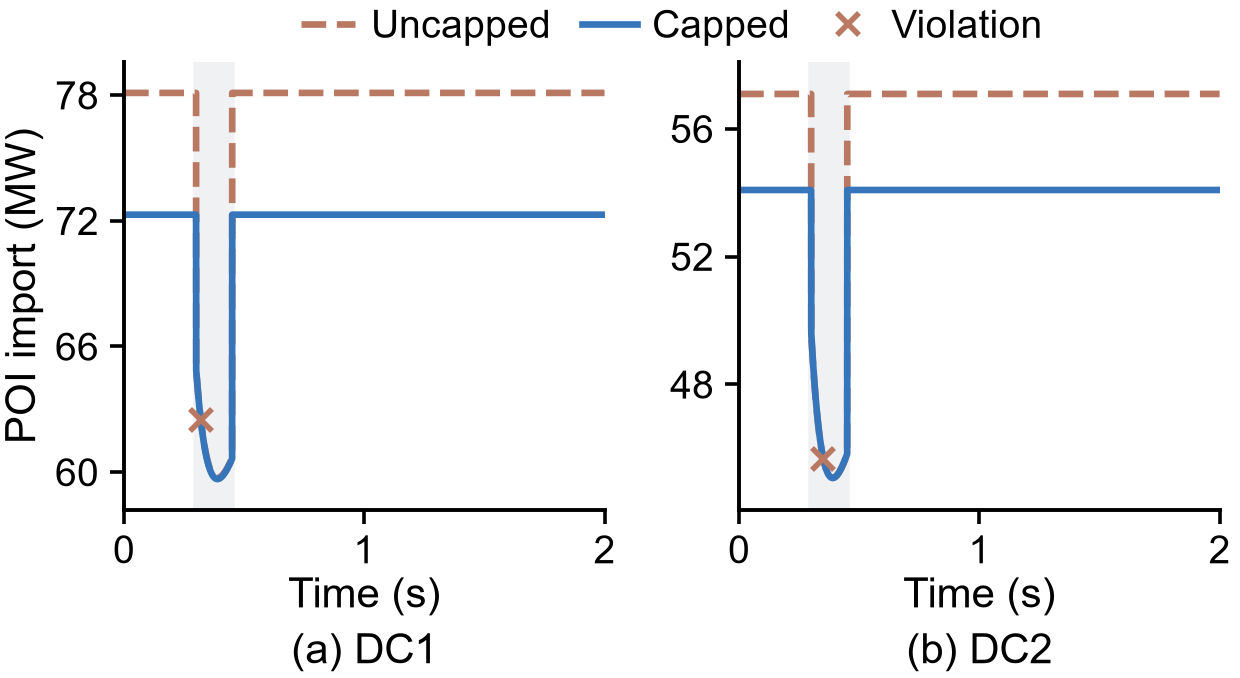}
\caption{Grid-side active-power imports under fault F1.}
\label{fig:fig5_p_grid_F1}
\end{figure}

The resulting IT-bus voltages and grid-side power imports are shown in Figs.~\ref{fig:fig4_v_it_F1} and~\ref{fig:fig5_p_grid_F1}, respectively. Without power capping, the voltage sag reduces the grid-side UPS power capability, resulting in a transient power deficit that must be compensated by the BESS and DC-link energy storage. The maximum reductions in grid-side power import reach 23.54\% for DC1 and 21.06\% for DC2, both exceeding the prescribed 20\% limit. Meanwhile, the minimum IT-bus voltages decrease to 0.888~pu and 0.922~pu, respectively, causing DC1 to violate the internal voltage ride-through requirement.

With power capping, the reduced pre-fault demand decreases the transient power deficit during the fault period. In the illustrated case, both IT-bus voltages remain above 0.99~pu, while the maximum reductions in grid-side power import decrease to 17.47\% and 16.68\%, respectively. The final scheduling solution satisfies the ride-through requirements for both data centers under all considered operating periods and fault scenarios. These results demonstrate that satisfying steady-state network constraints alone does not guarantee transient ride-through capability, and that incorporating VRT constraints enables coordinated data-center power capping to improve fault-period feasibility.

\section{Conclusion}
This paper proposes a VRT-aware data-center power-capping framework integrating ACOPF scheduling with dynamic ride-through assessment. Results demonstrate that steady-state feasible operating points may violate transient VRT requirements, while dynamic-constrained power capping effectively restores ride-through security. The study highlights the necessity of dynamic security assessment for determining operational limits of large-scale data centers.

\bibliographystyle{IEEEtran}
\bibliography{ref}
\end{document}